\documentclass[12pt,preprint]{aastex}

\shorttitle{Modest Cooling Flows}
\shortauthors{Bregman et al.}

\begin{document}


\title{\ion{O}{6} Observations of Galaxy Clusters: Evidence for Modest Cooling Flows}

\author{Joel N. Bregman\altaffilmark{1}, A.C. Fabian\altaffilmark{2},  
Eric D. Miller\altaffilmark{3}, and Jimmy A. Irwin\altaffilmark{1}}

\email{jbregman@umich.edu}

\altaffiltext{1}{Department of Astronomy, University of Michigan, Ann Arbor, MI 48109}
\altaffiltext{2}{Institute of Astronomy, Madingley Road, Cambridge CB3 0HA}
\altaffiltext{3}{Kavli Institute for Astrophysics and Space Science, MIT, Cambridge, MA 02139}

\begin{abstract}

A prediction of the galaxy cluster cooling flow model is that as gas cools
from the ambient cluster temperature, emission lines are produced in gas
at subsequently decreasing temperatures.  Gas passing through
10$^{5.5}$ K emits in the lines of \ion{O}{6} $\lambda \lambda$1032,
1035, and here we report a {\it FUSE\/} study of these lines in three
cooling flow clusters, Abell 426, Abell 1795, and AWM 7.  No emission
was detected from AWM 7, but \ion{O}{6} is detected from the centers of Abell 426
and Abell 1795, and possibly to the south of the center in Abell 1795, where X-ray and
optical emission line filaments lie.  In Abell 426, these line luminosities
imply a cooling rate of 32$\pm$6 M$_{{\rm \odot}{}}$ yr$^{-1}$ within the
central r = 6.2 kpc region, while for Abell 1795, the central cooling rate is
26$\pm$7 M$_{{\rm \odot}{}}$ yr$^{-1}$ (within r = 22 kpc), and about 42$\pm$9
M$_{{\rm \odot}{}}$ yr$^{-1}$ including the southern pointing.

Including other studies, three of six clusters have \ion{O}{6} emission,
and they also have star formation as well as emission lines from 10$^4$ K
gas.  These observations are generally consistent with the cooling flow
model but at a rate closer to 30 M$_{{\rm \odot}{}}$ yr$^{-1}$ than
originally suggested values of 10$^2$--10$^3$ M$_{{\rm \odot}{}}$
yr$^{-1}$.

\end{abstract}


\section{Introduction}

Before the launch of {\it XMM-Newton\/} and {\it Chandra\/}, the X-ray data
provided strong support for the model of cooling flows \citep{fab94}.  The X-ray
imaging data of galaxy clusters showed a temperature decrease into the
centers of these systems, typically by a factor of two from the ambient
temperature of about 4-10 keV (5-12$\times$10$^{{\rm 7}}$ K).  Within the inner 100 kpc,
the radiative cooling times for the hot gas is shorter than a
Hubble time, leading to the apparent inevitability that the gas would
radiatively cool to much lower temperatures, and this is the basis of the
cooling flow model with cooling rates of 30-1000 M$_{{\rm \odot}{}}$ yr$^{{\rm -1}}$.  However, the
existence of cooling flows has often been questioned by observers in
other wavebands, who find some evidence for cooled gas and star
formation, but at a level far below the rate inferred from the 
X-ray data \citep{mcnam89,mcnam93,card98}.

Consequently, it was anticipated that {\it XMM\/} grating spectra would show
strong emission from gas that was well below the ambient temperature of
the medium, representing the substantial amounts of cooling gas.  While
the {\it XMM\/} data showed the expected luminosities of the high ionization
line, such as \ion{Fe}{21} to \ion{Fe}{24}, the emission from the lines that trace 
cooler gas were missing (e.g., \ion{Fe}{17}, \ion{O}{7}; \citealt{pete03}).  
According to these line ratios, the gas was not obviously cooling
below about 2 keV ($2.3 \times 10^{7}$ K).  Either some heating mechanism turns on
at 2 keV, preventing the gas from cooling further, or the gas continues to
cool but these lines are not produced, due to metal inhomogenieties,
turbulent mixing with cooler gas, or heating by AGNs \citep{fab01a}.

There is still good evidence for significantly cooler gas (e.g., 10$^{4}$ K gas in Perseus), which
may be the end product of the cooling flow \citep{cowi83,hu83,hu85,ferr97, cons01}.  
However, unless one can find gas at temperatures between
10$^{4}$ K and 10$^{7}$ K, we would have to consider an alternative picture for
the cooler gas, where it is mass lost from galaxies \citep{sparks97}.  A particularly good tracer of
cooling gas is the \ion{O}{6} $\lambda \lambda$ 1032, 1038 doublet, because it
dominates the radiative cooling function where it is present ($\sim 10^{5.5}$ K), it
cannot be produced by photoionization by stars, and it lies above 912 \AA, 
so it is accessible with the {\it Far Ultraviolet Spectroscopic Explorer\/}
({\it FUSE\/}).  This line has been detected in Abell 2597 by \citet{oeger01a}
where they report a cooling rate of about 40 M$_{\odot}$ yr$^{-1}$ within
the inner 36 kpc of the cluster.  They also report a nondetection, for Abell
1795, and \citet{lecav04} report nondetections for
Abell 2029 and Abell 3112.  Here we report on {\it FUSE\/} observations of
Abell 426 (the Perseus cluster), Abell 1795 (off-center observations), and
AWM 7, all clusters that are bright in X-rays and originally expected to
contain cooling flows.  We find evidence for \ion{O}{6} emission from Perseus
and Abell 1795 and discuss the importance of these observations in light
of X-ray and other related observations.  We use a Hubble constant of
70 km s$^{-1}$ Mpc$^{-1}$.

\section{Observations}

All of the {\it FUSE\/} observations were obtained with the large aperture (30\arcsec\
square; {\it LWRS\/}) in time-tag mode.  The observations were screened for
particle outbursts, with the resulting good data separated into day and
night components.  
Each individual detector segment for each observation was combined using 
the procedure TTAG\_COMBINE before applying the CALFUSE pipeline (origianally,
version 2.4 was used but we also tried version 3.0, which led to no 
noticable changes for the spectral region of interest). 
For observations obtained during the day, the
background is significantly higher than for observations taken at night. 
The background is composed of a combination of discrete airglow lines
(e.g., from H$\beta$, \ion{O}{1}, N$_2$) plus a diffuse scattered component.  For the
data taken during the day, this background is often stronger than the
weak signals that we seek, so we usually used the data obtained during
nighttime (discussed separately for each of our targets).  For the Perseus
cluster and AWM 7, the redshifted \ion{O}{6} doublet falls near 1050 \AA, and
in this regions the channel with the highest S/N is the LiF1a.  As the
addition of other channels reduces the S/N, we discuss only the LiF1a data
for these clusters.  For Abell 1795, the \ion{O}{6} line is shifted into the 1095--1098 \AA\ 
region, where the only useful detector is the LiF2a channel.  For
purposes of displaying the data, the spectra are binned and smoothed,
although for the extraction of line fluxes, the unsmoothed data were used.

\subsection{Abell 426 (Perseus Cluster)}

\begin{figure}
\plotone{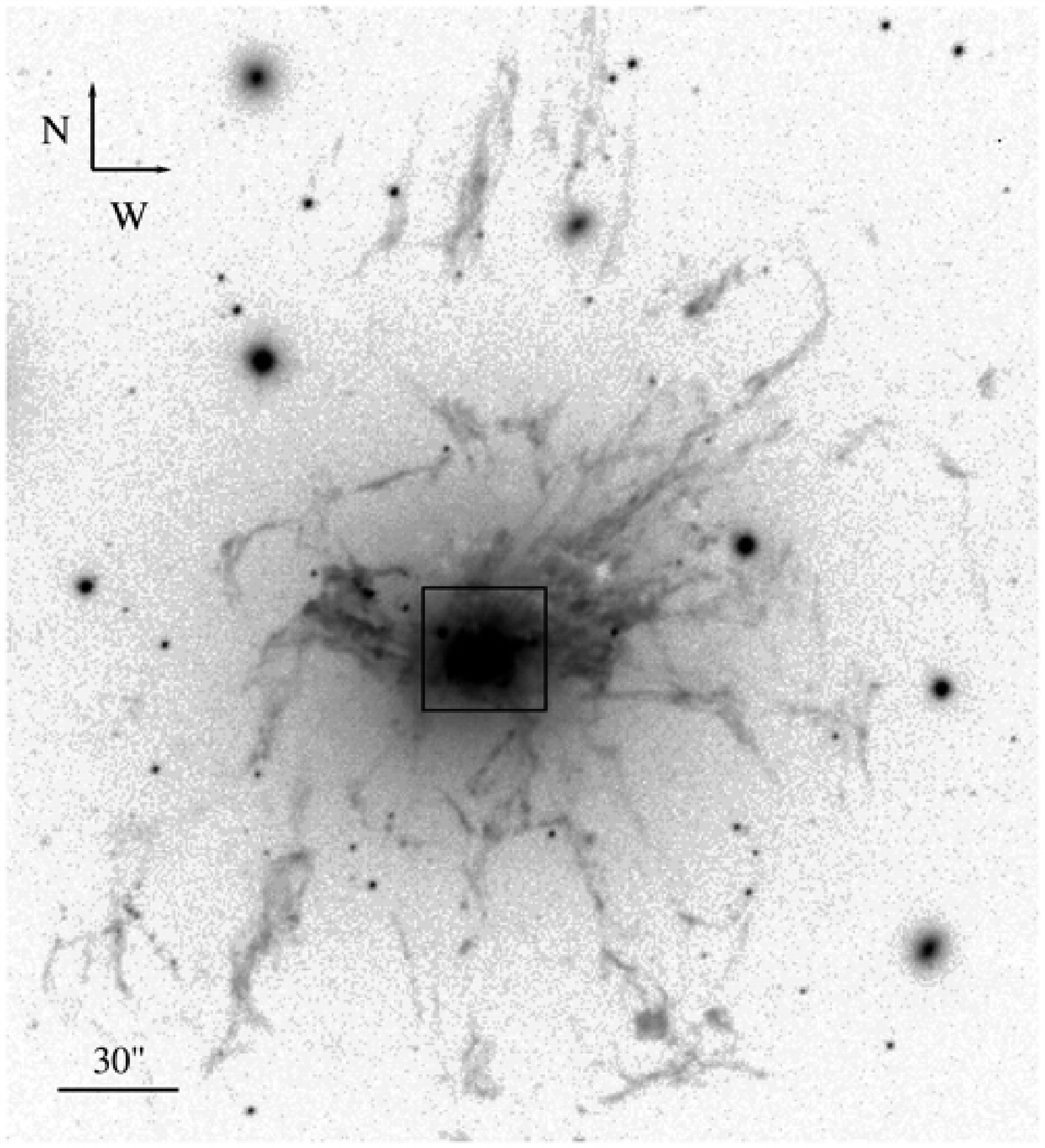}
\caption{The {\em FUSE\/} LWRS aperture is superimposed upon the
H$\alpha$ image of \citet{cons01}, which shows the low velocity
emission (5200 km s$^{-1}$) along with the continuum of NGC 1275 in
Abell 426.  The {\em FUSE\/} aperture only includes part of the
extensive optical emission line gas.}
\label{fig:A426_Halpha}
\end{figure}

This is one of the most frequently observed clusters and it was one of the
most compelling cases for cooling flows, as it showed an X-ray surface
brightness distribution with a sharply peaked diffuse emission, extensive
filaments from 10$^{4}$ K gas, and ongoing star formation \citep{norg90,
mcnam96,sand04}.  It also hosts an
AGN with strong extended radio emission (3C 84) and the central object
has been listed as a Seyfert 1.5 galaxy as well as a BL Lac type object.  In
addition to the central dominant galaxy, NGC 1275, at a redshift of 5264
km s$^{-1}$, there is a higher velocity gaseous component, at about 8200 km
s$^{-1}$ (first noted by \citealt{mink55}), which contains warm ionized gas, neutral gas,
and molecular gas \citep{jaffe90,don00,krab00}.  The emission line gas is seen over a 6\arcmin\
diameter region, which is similar in size to the cooling radius of the X-ray
emission \citep{cons01}.  Consequently, only a fraction of the cooling gas may be
contained in the LWRS (30\arcsec\ square centered on NGC 1275; Figure \ref{fig:A426_Halpha}), 
so we may be missing a considerable amount of cooling gas, although this
pointing contains the brightest H$\alpha$ emitting gas and X-ray emitting
gas.  The data were taken in 16 October 2002, with a total exposure time
of 29.1 ksec, of which 13.0 ksec is night data, used in our analysis.

\begin{figure}
\plotone{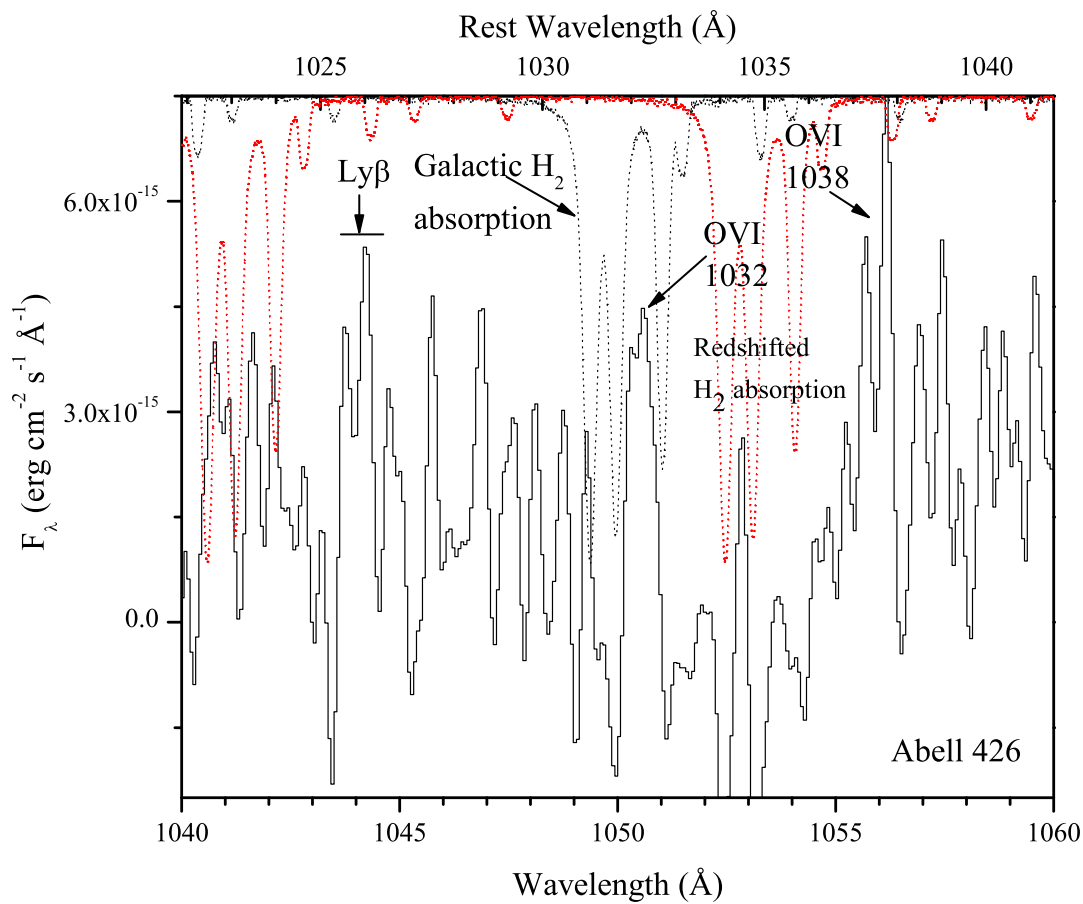}
\caption{The {\em FUSE\/} spectrum of the center of Abell 426 shows
both \ion{O}{6} emission lines plus a faint optical continuum and a
Ly$\beta$ line.  The rest wavelength is given at the top and the observed
wavelength at the bottom.  Galactic H$_2$ absorption (dotted black line) has
probably absorbed some of the strong \ion{O}{6} line but the weaker line
is unaffected.  H$_2$ absorption within Abell 426 (dotted red line;
8200 km s$^{-1}$) is present but does not absorb the either \ion{O}{6} line.
}
\label{fig:A426_OVI}
\end{figure}

Several emission lines are detected, including Ly$\beta$ and both \ion{O}{6} lines
near the redshift of NGC 1275 (Figure \ref{fig:A426_OVI}).  The analysis of these lines is complicated
by a variety of absorption lines, mostly from the Galaxy, but also from
material within the Perseus cluster.  From the Milky Way, we see strong
absorption lines from H$_2$ in addition to the usual atomic lines, but this is
to be expected given the low Galactic latitude and high extinction in this
direction (A$_B$ = 0.70 mag; \citealt{schleg98}).  This absorption affects the strong \ion{O}{6}
$\lambda$ 1032 line, which is absorbed on its blue and red sides by Galactic
H$_2$.  In addition, there is absorption due to H$_2$ from the higher redshift
system in Perseus (8200 km s$^{-1}$), but this is to be expected as molecular
gas was previously known to exist at this velocity \citep{jaffe90}.  Fortunately, no strong
absorption features occur near the weaker \ion{O}{6} line, so we are able to
obtain a line flux.  For the \ion{O}{6} $\lambda$ 1038 line, the flux is 
2.9 $\times$ 10$^{-15}$ erg cm$^{-2}$ s$^{-1}$ ($\pm$ 0.4 $\times$ 10$^{-15}$ erg cm$^{-2}$ s$^{-1}$), 
the line center is at 1055.9 $\pm$ 0.1 \AA, and
the width is 0.7 $\pm$ 0.2 \AA\ (FWHM), while for the 1032 \AA\ line, the
flux is  2.9 $\times$ 10$^{-15}$ erg cm$^{-2}$ s$^{-1}$ , which we regard as a lower limit
due to the absorption.  It is difficult to correct the strong line for
absorption in a reliable manner, so we use the line ratio (of 2:1) and the
strength of the $\lambda$ 1038 line to infer the strength of the stronger line
to be about 5.8 $\times$ 10$^{-15}$ erg cm$^{-2}$ s$^{-1}$ ($\pm$ 0.9 $\times$ 10$^{-15}$ erg cm$^{-2}$ s$^{-1}$).  
After making an extinction
correction, F($\lambda$ 1032) = 4.2 $\times$ 10-14 erg cm$^{-2}$ s$^{-1}$
($\pm$ 0.6 $\times$ 10$^{-14}$ erg cm$^{-2}$ s$^{-1}$), and for a
distance of 75.2 Mpc, the luminosity in the $\lambda$ 1032 line is 2.9 $\times$
10$^{40}$ erg s$^{-1}$ ($\pm$ 0.4 $\times$ 10$^{40}$ erg s$^{-1}$.  
In addition, the Ly$\beta$ line is detected in emission, which
was expected since Ly$\alpha$ was detected previously.

\subsection{Abell 1795}

The cluster has a mean redshift z = 0.062476 $\pm$ 0.00030
\citep{oeger01b} and the velocity dispersion of the cluster is 810
km s$^{-1}$ ($\bigtriangleup$z = 0.0027).  The velocity of the central dominant galaxy is slightly
higher, at z = 0.06326.  This cluster has a filament that extends about 40\arcsec\ 
to the south of the central dominant galaxy and is detected in both
optical emission line gas as well as diffuse X-rays \citep{hu85,fab01b}.  The filament
is at least a factor of two cooler than the ambient cluster gas and the
radiative cooling time is 3 $\times$ 10$^8$ yr, about the age inferred from its
length and velocity.  We obtained an observation at this off-center
location with the LWRS (30\arcsec\ south of the CD galaxy), and in addition, we reanalyzed the central
pointing (originally discussed by \citealt{oeger01a}), which contains the central
dominant galaxy, the peak of the X-ray emission, and a significant
fraction of the 10$^4$ K gas.  The total exposure time of the central pointing
is 37.1 ksec, of which 17.0 ksec were useful nighttime data, and those are
the observations used below.

\begin{figure}
\plotone{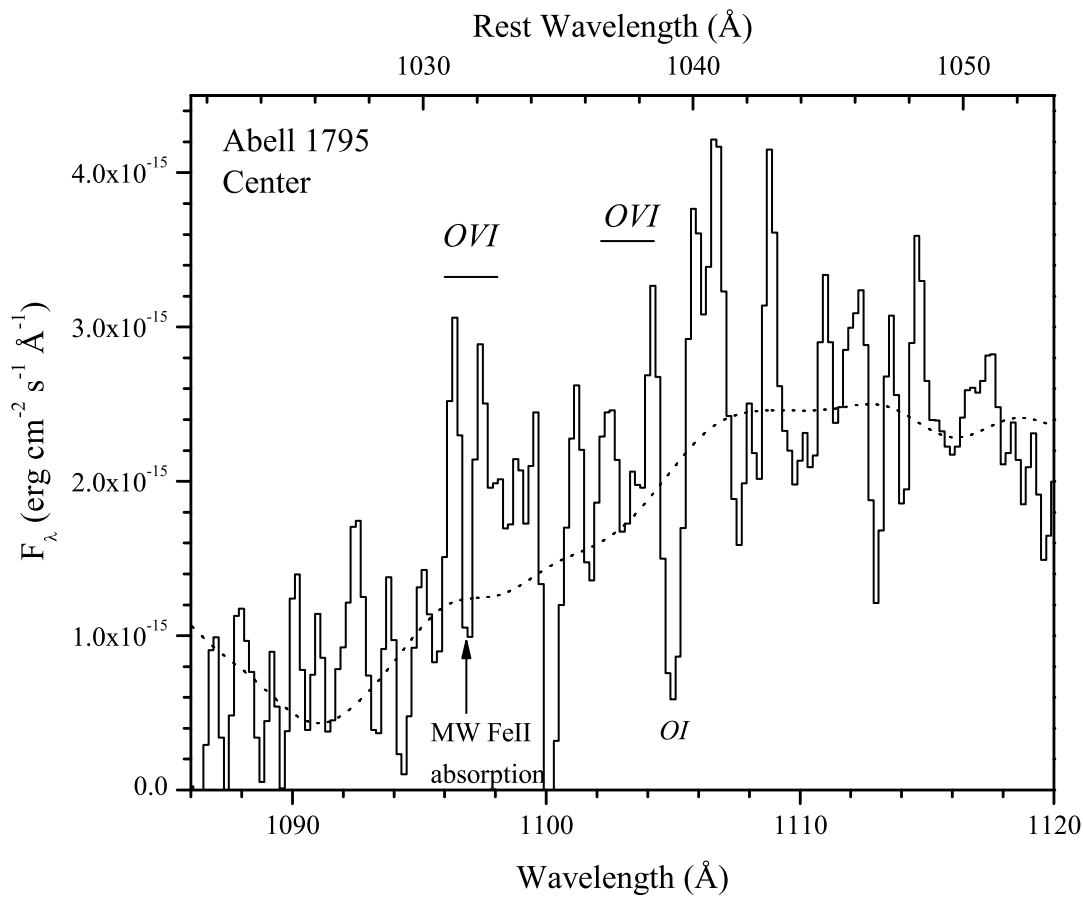}
\caption{The {\em FUSE\/} LWRS spectrum of the center of Abell 1795,
along with the continuum of the central galaxy of the Fornax cluster,
NGC 1399 (dotted line), which has no emission lines (same wavelength 
scales as above).  The strong
\ion{O}{6} emission line lies above the continuum and is probably partly
absorbed by a strong Galactic Fe II line at 1096.88 \AA.  The only other
significant feature is emission at 1107 \AA. }
\label{fig:A1795_OVI_n1399}
\end{figure}

The central galaxy has an apparent magnitude of 15.2, so if the ratio of V
to FUV flux was typical for an early-type galaxy (e.g., NGC 5846), the
flux near 1115 \AA\ (1050 \AA\ in the rest frame) would have been about 5
$\times$ 10$^{-16}$ erg cm$^{-2}$ s$^{-1}$ (after \citealt{breg05}), 
but the observed flux is 2.7 $\times$ 10$^{-15}$ erg
cm$^{-2}$ s$^{-1}$, about five times larger (Figure \ref{fig:A1795_OVI_n1399}).  
There is a range in the FUV to V flux of early-type galaxies
\citep{burst88}, and some luminous galaxies are a magnitude brighter in
their FUV flux (e.g., NGC 1399), so it is possible that starlight from an
old population can account for half of the FUV continua.  The
remainder, if not most of the FUV continua, is probably due to young
stars, as discussed by \citet{mitt01} in their imaging study of the
center of Abell 1795 with the optical monitor on {\it XMM\/}, which has UV
channels.

\begin{figure}
\plotone{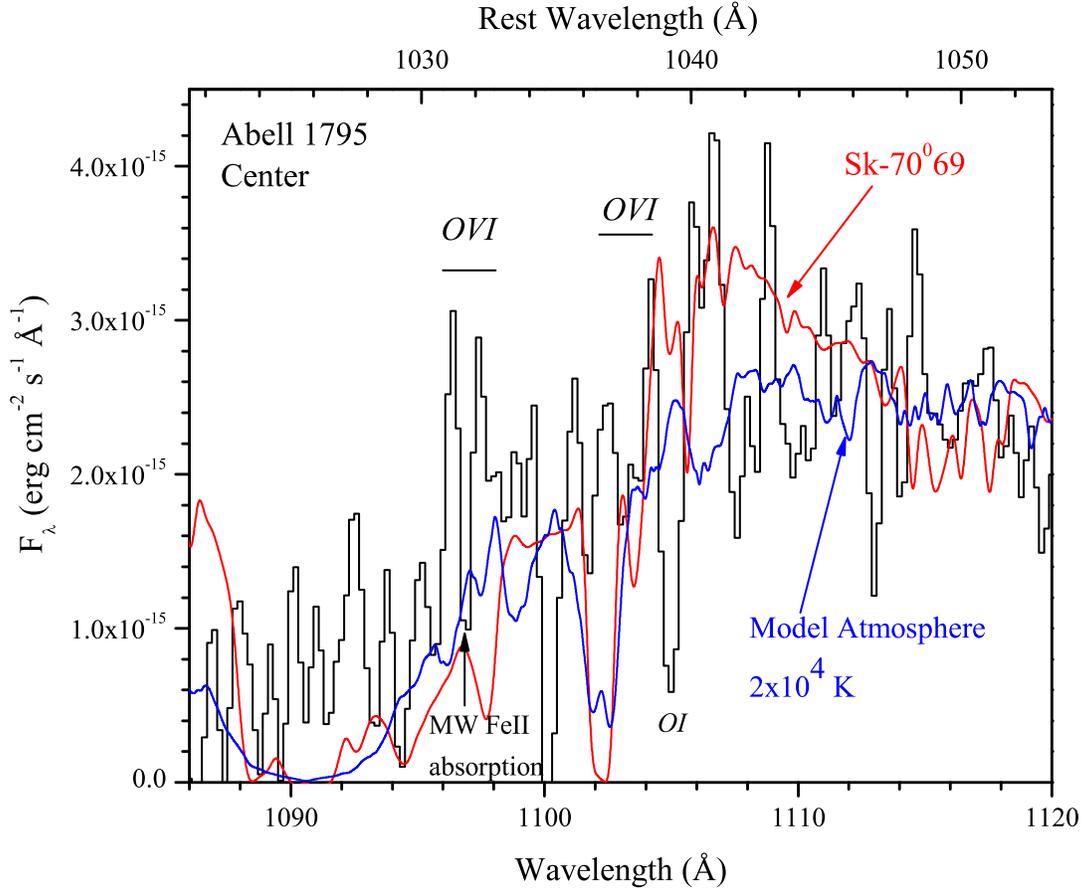}
\caption{The same spectrum as above, except with the continuum of
NGC 1399 replaced with the continuum of two hot stars, Sk-70$\arcdeg$69 
(red line; an O5V star showing a strong wind;
\citealt{walb02}), and a $2 \times 10^4$ K model atmosphere (log$g$ =
4; solar metallicity; blue line; we use the same wavelength 
scales as above).  The star Sk-70$\arcdeg$69 rises up to
1107 \AA\ (1041 \AA\ in the rest frame), which coincides with the peak
in the observed spectrum from Abell 1795.}
\label{fig:A1795_hotstars}
\end{figure}

When we scale the UV continuum of NGC 1399 to the region beyond 1110 \AA\ 
(1044 \AA\ in the rest frame), the two most significant features in the 
entire LiF2a spectrum occur at 1096.8 \AA\ and 1106.4 \AA.  This first 
feature corresponds to the location of the strong \ion{O}{6} line at 
the redshift of the central galaxy and the optical emission line gas in
Abell 1795.  We believe that the second feature is associated with young
hot stars that have been formed recently and create much of the UV continuum.
There can be a few significant spectral differences for hot stars in this
wavelength range \citep{pell02,walb02}.
The spectrum rises from the minimum caused by Ly$\beta$ absorption (1091 \AA),
although the feature is typically 10\AA\ FWHM and damped, leading to the
slow rise in both the young hot stars (Figure \ref{fig:A1795_hotstars}) and the old hot stars in
NGC 1399 (Figure \ref{fig:A1795_OVI_n1399}).  
Young hot stars can have a wind, and depending upon the mass flux,
P-Cygni profiles can occur \citep{walb02}, which causes the peak near 1104.5--1109.5 \AA\ 
(1038.5--1043.5 \AA\ in the rest frame), corresponding to the peak observed
in the LiF2a spectrum.  The weak \ion{O}{6} line is not
apparent, but its wavelength corresponds to an absorption feature in an
O/B star.  If we subtract the hot star stellar atmosphere from the spectrum,  
an emission line would appear at the location of the \ion{O}{6} $\lambda$ 1038 line,
so it is possible that both lines exist.

In trying to determine the strength of the strong \ion{O}{6} line, we note
that a strong Galactic \ion{Fe}{2} absorption line would lie near the 
line center.  Using the stellar continua as a baseline, we find that the
line flux is 2.3 $\times$ 10$^{-15}$ erg cm$^{-2}$ s$^{-1}$ ($\pm$ 
0.6 $\times$ 10$^{-15}$ erg cm$^{-2}$ s$^{-1}$ ), with a line center at 1069.9 (z =
0.0630) and a line width of 2.3 \AA\ FWHM.  
The absorption-corrected flux for the strong \ion{O}{6} line is 2.7
$\times$ 10$^{-15}$ erg cm$^{-2}$ s$^{-1}$ ($\pm$ 0.7 $\times$ 10$^{-15}$ erg cm$^{-2}$ s$^{-1}$ ), 
and for a distance of 268 Mpc, the luminosity is 
2.3 $\times$ 10$^{40}$ erg s$^{-1}$ ($\pm$ 0.6 $\times$ 10$^{40}$ erg s$^{-1}$).

The \ion{O}{6} line emission from the central pointing was listed as an upper
limit by \citet{oeger01a}.  Since that time, improvements in the pipeline
processing have reduced the noise in the spectral extraction and our understanding 
of the underlying stellar continuum has improved.  There was also a
concern of significant airglow contamination with the strong line from
N$_2$ when pointing near the illuminated disk of the Earth, the strongest
near the \ion{O}{6} line being at 1098.1 \AA\ (there is a much 
stronger N$_2$ line at 1034 \AA).  In comparing the night-only data with the
day-plus-night data, there is a tremendous reduction in the 1034 \AA\ line,
and a small fraction of the emission at 1098.1 \AA\ decreases.  Therefore, it is unlikely 
that the 1098.1 \AA\ airglow line contributes significantly to the line, especially as it is 
to the red of the line defined by its FWHM.

\begin{figure}
\plotone{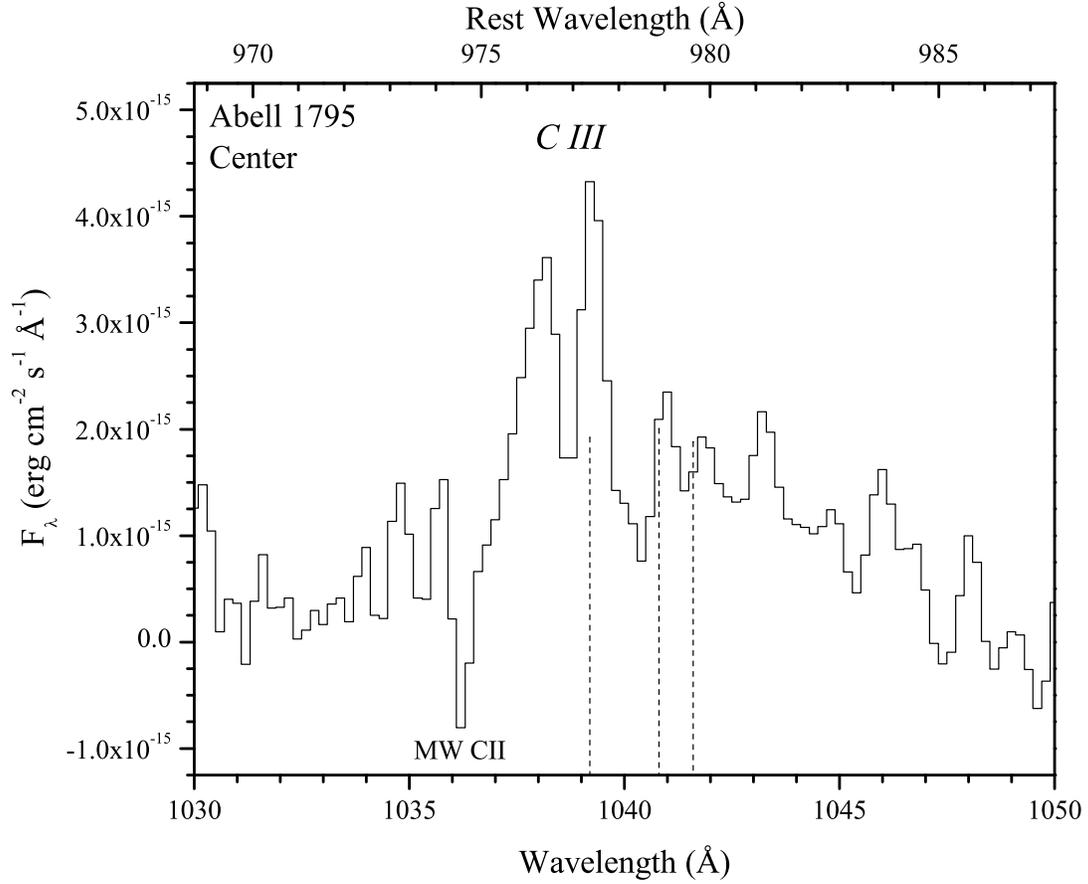}
\caption{The spectral region containing the redshifted \ion{C}{3} line 
in the central pointing of Abell 1795,
along with markers of the location for airglow lines (vertical dashed
lines), where the height of the dashed lines signifies the line relative strength.  The
red side of the \ion{C}{3} line may have a small amount of contamination
due to an airglow line.}
\label{fig:A1795_CIII}
\end{figure}

Although it is not central to the primary science, the \ion{C}{3} $\lambda$ 977 line is
detected in the LiF1a channel (Figure \ref{fig:A1795_CIII}).  
It is the strongest feature in that channel not due 
to airglow and has a central wavelength of 1038.5 \AA, with a flux of
4.2$\times$ 10$^{-15}$ erg cm$^{-2}$ s$^{-1}$ $\pm$ 
0.5 $\times$ 10$^{-15}$ erg cm$^{-2}$ s$^{-1}$.

The observation for the off-center pointing of Abell 1795 was taken at
two times in 2004, with total exposure times of 6.49 ksec and 18.86 ksec,
although we are using only the night data since the background is
significantly lower, leading to a higher S/N for the final spectra.  The final
useful night time data had exposure times of 4.48 ksec and 12.7 ksec or a
total time of 17.2 ksec.  We use the combined night spectrum for the
analysis, in which the airglow lines from oxygen and nitrogen are so weak
as to be undetectable.

\begin{figure}
\plotone{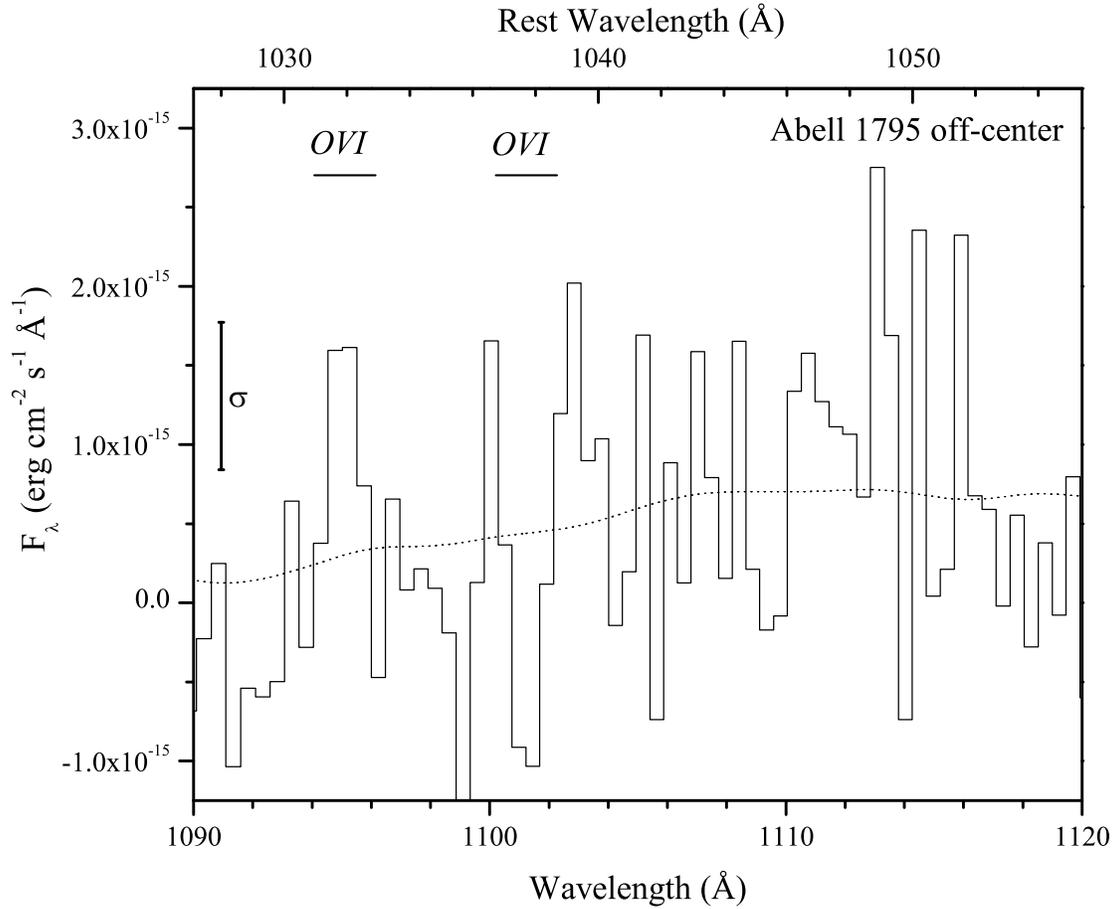}
\caption{The {\em FUSE\/} LWRS spectrum taken 30\arcsec\ from the
center of Abell 1795, along with the continuum of NGC 1399 (dotted
line).  The strongest feature in this LiF2a spectrum is that of the strong
\ion{O}{6} line, about a 3$\sigma$ feature; the weaker \ion{O}{6} line
is not detected.}
\label{fig:A1795off}
\end{figure}

The detector with the highest S/N in the wavelength region where the
redshifted emission would occur is the LiF2a channel, and the strongest
feature in this channel occurs at 1095.1 \AA, which is approximately the
redshifted wavelength of the \ion{O}{6} $\lambda$ 1032 line, 
z = 0.06122 (Figure \ref{fig:A1795off}).  This line is
370 km s$^{-1}$ to the blue of the mean redshift of the cluster, which has a velocity
dispersion of 810 km s$^{-1}$ \citep{oeger01b}.  However, the optical emission
line gas is also blueshifted at this location by 100-200 km s$^{-1}$, and the
line width is broad and extends to the velocity of the \ion{O}{6} emission \citep{hu85}.
This line is detected at the 3
$\sigma$ level and with a flux of 1.4$\times$10$^{-15}$ erg cm$^{-2}$ s$^{-1}$ $\pm$ 
0.5$\times$10$^{-15}$ erg cm$^{-2}$ s$^{-1}$, and the FWHM is 1.0 $\pm$ 0.2  \AA.  The
weaker \ion{O}{6} line, which is not detected, would have been a 1.5 $\sigma$
feature.  To be conservative, we view this as a possible detection and
discuss it accordingly.  A very faint stellar continuum is present as well,
with a peak emission in the 1105-1120 \AA\ range (1040-1053 \AA\ in the
rest frame), as expected for a hot stellar continuum from young stars;
the zero point is too uncertain to place a secure flux level on the continuum. 
Another feature that one might expect, either from cooling gas or
photoionized gas, is the \ion{C}{3} $\lambda$ 977 line, and there is a marginal
detection of this line at the 2 $\sigma$ level, 1.4$\times$10$^{-15}$ erg cm$^{-2}$ s$^{-1}$
$\pm$ 0.7$\times$10$^{-15}$ erg cm$^{-2}$ s$^{-1}$.

\subsection{AWM 7}

The cluster AWM7 is a nearby X-ray bright cluster with an elliptical galaxy at
the center, at a velocity of 5194 km s$^{-1}$.  It lies at l, b = 146\arcdeg\, -15.6\arcdeg\, 
and the extinction is E(B-V) = 0.116 mag \citep{schleg98}.  It was
observed on 26 September 2003 with a total exposure time of 21.1 ksec, but
the S/N is highest when using the night-only data, which comprises 14.1 ksec.

The spectra show no features at all, aside from the usual H$\beta$ airglow line.
From the noise characteristics of the spectrum, we derive a 3$\sigma$ upper limit of
3$\times$10$^{-15}$ erg cm$^{-2}$ s$^{-1}$ to an \ion{O}{6} line of 
width 300 km s$^{-1}$.  After correction for
extinction, the strong \ion{O}{6} line has an upper limit to its luminosity of
8 $\times$ 10$^{39}$ erg s$^{-1}$. 

\section{Interpretation and Discussion}

The OVI luminosities can be converted into mass cooling rates using radiative
cooling models.  Since these OVI lines carry the cooling through this temperature region,
the line strength is the product of the thermal energy of the gas at 10$^{5.5}$ K
and the rate at which matter is cooling through that region \citep{edgar86,voit94}.  Factors such as the
metallicity of the gas, departures from thermal equilibrium, and whether the gas
is isochoric or isobaric only modify the line strength modestly.
For the conversion from L($\lambda$1032) to a mass cooling rate, we use 
the value discussed by \citet{breg01} (see \citealt{edgar86} and \citealt{voit94}),
of $\dot{M}$ = (L($\lambda$1032)/9$\times 10^{38}$ erg s$^{-1}$) M$_{\odot}$ yr$^{-1}$.
Using this conversion, the upper limit to the cooling rate for AWM 7 is
9 M$_{{\rm \odot}{}}$ yr$^{-1}$ within r = 6.1 kpc, and the detected cooling rate 
for Abell 426 is 32$\pm$5 M$_{{\rm \odot}{}}$ yr$^{-1}$ within the central r = 6.2 kpc.  
For Abell 426, the cooling radius is about an order of magnitude larger than the 
{\it FUSE} aperture and the optical emission line gas region is also considerably
larger, so we may be missing much of the \ion{O}{6} emission.  Therefore, we regard
this cooling rate as a minimum value.

The cooling rate for the central pointing of Abell 1795 is about 26$\pm$7 M$_{{\rm \odot}{}}$ yr$^{-1}$
in the central r = 22 kpc region.  If we treat the off-center pointing as a 
detection, it represents an additional 16$\pm$6 M$_{{\rm \odot}{}}$ yr$^{-1}$, or a
total measured in the cluster of 42$\pm$9 M$_{{\rm \odot}{}}$ yr$^{-1}$.  Once again, 
much of the cooling gas may not be contained within our apertures, so the total
cooling rate for the cluster is likely to be larger.

We can compare our measurements with the three other rich clusters that have been 
observed with {\it FUSE}, Abell 2597, Abell 2029, and Abell 3112 \citep{oeger01a,lecav04}.
When we convert these published values to the same Hubble constant and cooling
rate scale, the upper limits for Abell 2029 and Abell 3112 are 27 M$_{{\rm \odot}{}}$ yr$^{-1}$ 
(r = 27 kpc), and 25 M$_{{\rm \odot}{}}$ yr$^{-1}$ (r = 26 kpc), while for Abell 2597, the
detected value is 35 M$_{{\rm \odot}{}}$ yr$^{-1}$ (r = 30 kpc).  In summary, of the 
six clusters observed, three are detected in the \ion{O}{6} lines.

There are other characteristics that distingish the clusters detected in \ion{O}{6}
from those that were not.  The three clusters, Abell 426, Abell 1795, and Abell 2597 
have extensive optical emission lines and also have evidence for recent or ongoing
star formation.  Although the other galaxy clusters were considered cooling flow clusters
from their broad-band X-ray data, they are not known to have optical emission line filaments
or star formation, such as Abell 2029 \citep{hu85,mcnam89,mcnam93,card98}.  
It had been suggested by others that the combination
of peaked X-ray emission with short cooling times, optical emission line gas, and young
stars constituted a strong case for the cooling flow paradigm (e.g., \citealt{card98}).  The
counterargument was that the cooled gas was mass loss from a recently stripped
galaxy, and the disturbance had led to star formation in the gas \citep{sparks97}.  In this
view, the gas does not cool from the hot ambient cluster medium.  However, the
presence of \ion{O}{6} emission shows that, at least in some clusters, gas is 
cooling through the 10$^{5.5}$ K regime at a rate of at least 30 M$_{{\rm \odot}{}}$ yr$^{-1}$.
The original cooling flow estimates for these clusters are typically an order
of magnitude larger.  Either the \ion{O}{6} emission that we detected
is only a fraction of the cooling gas, which is possible since the apertures
cover only a fraction of the cooling flow region, or the cooling rate is an order of
magnitude lower than originally estimated.  A significantly lower cooling rate
would, in general, be consistent with the weak X-ray cooling lines, and models
have been developed along these lines \citep{soker03}.

We can compare the cooling rates inferred from the \ion{O}{6} emission with
the constraints from the X-ray data, and in particular, the {\it XMM\/} RGS data.
There is not yet a published RGS spectrum for Abell 426, but there are analyses for
both Abell 1795 and Abell 2597.  The EPIC plus RGS spectrum for Abell 2597 implies an
X-ray cooling rate of 50$\pm$10 M$_{{\rm \odot}{}}$ yr$^{-1}$ 
(adjusted for the H$_0$ used here) and a cooling radius of about 90 kpc \citep{morr05}.
This cooling rate is somewhat greater than the value that we measure, 
suggesting that cooling gas may extend beyond the {\it FUSE\/} LWRS
aperture.  If the cooling rate increases linearly with radius out to 
the cooling radius, as suggested by theory \citep{fab94}, the value 
inferred from the {\it FUSE\/} data would become 50-60 M$_{{\rm \odot}{}}$ yr$^{-1}$,
the same as found from the X-ray data. 
The study that includes Abell 1795 \citep{pete03}
places an upper limit to the cooling of 30 M$_{{\rm \odot}{}}$ yr$^{-1}$, which 
may be in conflict with the \ion{O}{6} data that implies 30--40 M$_{{\rm \odot}{}}$ yr$^{-1}$.
If such a conflict exists, it would imply that non-steady cooling of material
is involved, such as turbulent mixing layers \citep{slavin93}, although there
are other possibilities \citep{fab01a}.

This promising line of study of \ion{O}{6} emission is limited by the sensitivity of current instrumentation.
Natural follow-up observations, such as the mapping of \ion{O}{6} throughout the potential 
cooling region will have to wait for future missions.

\acknowledgements
We would like to thank the {\it FUSE\/} team for their assistance in the 
collection and reduction of these data.  
Also, we would like to thank Birgit Otte and Renato Dupke for their comments
and suggestions in writing this paper.
This research has made use of the NASA/IPAC Extragalactic Database (operated by JPL, Caltech),
the Multimission Archive at Space Telescope, and the NASA Astrophysics 
Data System, operated under contract with NASA.
We gratefully acknowledge support by NASA through grants NAG5-9021, 
NAG5-11483, G01-2089, GO1-2087, GO2-3114, and NAG5-10765.

\clearpage

\end{document}